\documentclass[8pt,apl,superscriptaddress,pdftex,dvipsnames,twocolumn]{revtex4-1}

\usepackage{amsmath,amsthm,amssymb,amsfonts}
\usepackage{parallel,accents}
\usepackage{stmaryrd}

\usepackage{amsxtra,amstext,latexsym,dsfont} 

\normalfont
\usepackage{bm}

\usepackage[symbol]{footmisc}
\usepackage{perpage}
\MakePerPage[2]{footnote}

\usepackage[pdftex]{graphicx}
\usepackage[dvipsnames]{xcolor}

\usepackage{pgf,tikz}
\usetikzlibrary{snakes}
\usetikzlibrary{backgrounds}
\usetikzlibrary{arrows}
\usetikzlibrary{patterns}
\usetikzlibrary{fit}
\usetikzlibrary{calc}

\theoremstyle{plain}      
\theoremstyle{plain}      
\theoremstyle{plain}      
\theoremstyle{plain}      
\theoremstyle{definition} 
\theoremstyle{definition} 
\theoremstyle{definition} 
\theoremstyle{plain} 
\theoremstyle{definition} 
\theoremstyle{plain} 
\theoremstyle{definition} 
\theoremstyle{definition} 
\theoremstyle{definition}

\newcommand\bb{\mathbb}

\newcommand\te{\text}

\newcommand\op{\operatorname}

\newcommand\as{^\ast}

\newcommand\E{\bb{E}}

\newcommand\lb{\lbrace}
\newcommand\lt{\left}

\newcommand\p{\bb{P}}           

\newcommand\pri{^\prime}

\newcommand\rb{\rbrace}
\newcommand\rt{\right}

\newcommand\al{\alpha}
\newcommand\be{\beta}

\newcommand\ome{\omega}

\begin{document}
\title{Percolation under Noise: \\
       Detecting Explosive Percolation Using the Second Largest Component}
\date{\today}
\author{Wes Viles*}
\author{Cedric E.~Ginestet*}
\thanks{The authors marked $*$ have equally contributed to the production of
  this article. CEG is the corresponding author.}
\author{Ariana Tang}
\author{Mark A.~Kramer}
\author{Eric D.~Kolaczyk}
\affiliation{Department of Mathematics and Statistics, Boston University}
\thanks{EDK and CEG were supported by a grant from the Air Force
Office for Scientific Research (AFOSR), whose grant number is
FA9550-12-1-0102. MAK was supported by a Career Award at the Scientific Interface
from the Burroughs Wellcome Fund. The authors thank Dr. Sydney Cash
for providing the brain voltage data analyzed in Figure 1.}
\begin{abstract}
   We consider the problem of distinguishing classical 
   (Erd\H{o}s-R\'{e}nyi) percolation from explosive (Achlioptas) percolation, under
   noise. A statistical model of percolation is constructed 
   allowing for the birth and death of edges as well as the presence
   of noise in the observations. This graph-valued stochastic process
   is composed of a latent and an observed non-stationary process,
   where the observed graph process is corrupted by Type I and Type II
   errors. This produces a hidden Markov graph model. We show that for
   certain choices of parameters controlling the noise, the
   classical (ER) percolation is visually indistinguishable from
   the explosive (Achlioptas) percolation model. In this setting, we
   compare two different criteria for
   discriminating between these two percolation models, based on a
   quantile difference (QD) of the first component's size and on the
   maximal size of the second largest component. We
   show through data simulations that this second criterion
   outperforms the QD of the first component's size, in terms of
   discriminatory power. The maximal size of the second component
   therefore provides a useful statistic for distinguishing
   between the ER and Achlioptas models of percolation, under
   physically motivated conditions for the birth and death of edges,
   and under noise. The potential
   application of the proposed criteria for percolation detection 
   in clinical neuroscience is also discussed.
\end{abstract}
\maketitle

\section{Introduction} 
Understanding the emergence of organized structure in dynamic networks
remains an active research area.  In the study of random networks,
percolation --the sudden emergence of a giant connected component
(GCC)-- is of critical importance from a theoretical, applied and
statistical perspective. Percolation in the Erd\H{o}s-R\'{e}nyi (ER) model
constitutes one of the first examples of a fully characterized
mathematical phase transition \citep{Erdos1960,Alon2004}.  While the ER model of
percolation is an example of a (second order) continuous phase
transition, recent efforts have focused on identifying the conditions
under which a random network process can yield a (first order)
discontinuous percolation \citep{Riordan2011}.

One of most popular attempts to model discontinuous percolation has
been the Achlioptas' process and its variants \citep{Achlioptas2009}.  The
Achlioptas' product rule (PR) slows down the growth of the GCC by
favoring the creation of edges between small connected components. 
Although this particular percolation model has been shown to be, in fact,
continuous and therefore of second order
\citep{Riordan2011,Riordan2012}; it nonetheless provides an
interesting alternative to the ER model. 
Achlioptas' processes have indeed generated a substantial amount of
theoretical work, whereby authors have explored related strategies for 
producing explosive percolation in random networks
\citep{Bohman2001,Beveridge2007,Krivelevich2010}. 
In addition, \citet{Riordan2011} have shown that genuine first-order phase
transitions can be realized by systematically adding, at every step of
the process, the edge that joins the two smallest components in the
\textit{entire} network. 

Interest in network percolation has been fueled by its relevance to
several application domains. In clinical neuroscience, for instance,
epileptic seizures have been associated
with the sudden emergence of coupled activity across the brain 
 \citep{Guye2006,Ponten2007,Schindler2007,Schindler2007a,
Kramer2010,Schindler2010}. The resulting functional networks --in which edges
indicate strong enough coupling between brain regions
\citep{Rubinov2010}-- are consistent with the notion of percolation. A
better understanding of
the type of phase transitions undergone at different stages of the
seizure, may aid in the development of novel strategies for the
treatment of epilepsy \citep{Kramer2012}.

The rich theory on percolation, and its application to real world
data, motivates the following question: How can we distinguish between
different percolation regimes in practice?
Previous theoretical work has concentrated on noise-free percolation,
which constitutes an idealized perspective on percolation
processes. In practice, however, the sampling of real-world networks
is likely to be corrupted by measurement errors.  Moreover, network
growth has generally been conceived as a monotonic process, whereby
only edge creations are allowed.  However, this assumption may be too
restrictive, since in real-world networks, the number of edges may
increase and decrease over time, in a stochastic manner (see example
in Figure 1).  Finally, to the best of our knowledge, there does not
currently exist a statistical framework for distinguishing between
different types of percolation regimes in the presence of edge birth
and death, and noise.

In this paper, we propose a framework to distinguish between different
percolation regimes in practice. To do so, we formulate the problem
of recognizing a percolation regime from noisy observations as
a question of statistical inference. Under this framework, we compare
the discriminatory power of two potential percolation features deduced
from the evolution of the first and second component of an observed
dynamic network.  We test this framework in simulation by constructing
a hidden Markov graph model, which encompasses both a non-stationary
latent process characterized by birth and death of edges, and an
observed graph process that introduces both Type I and Type II errors.
We show that edge death and noise make a statistic deduced from the
first component ineffective in distinguishing between the standard ER
second-order percolation and Achlioptas' explosive percolation.
However, a different detection criterion --based on the size of the
second component-- successfully discriminates between the two percolation
regimes in the presence of edge death and noise. These results
provide a framework for distinguishing percolation regimes in
practice.
\begin{figure}[t]
  \centering
 \includegraphics[width=8.5cm]{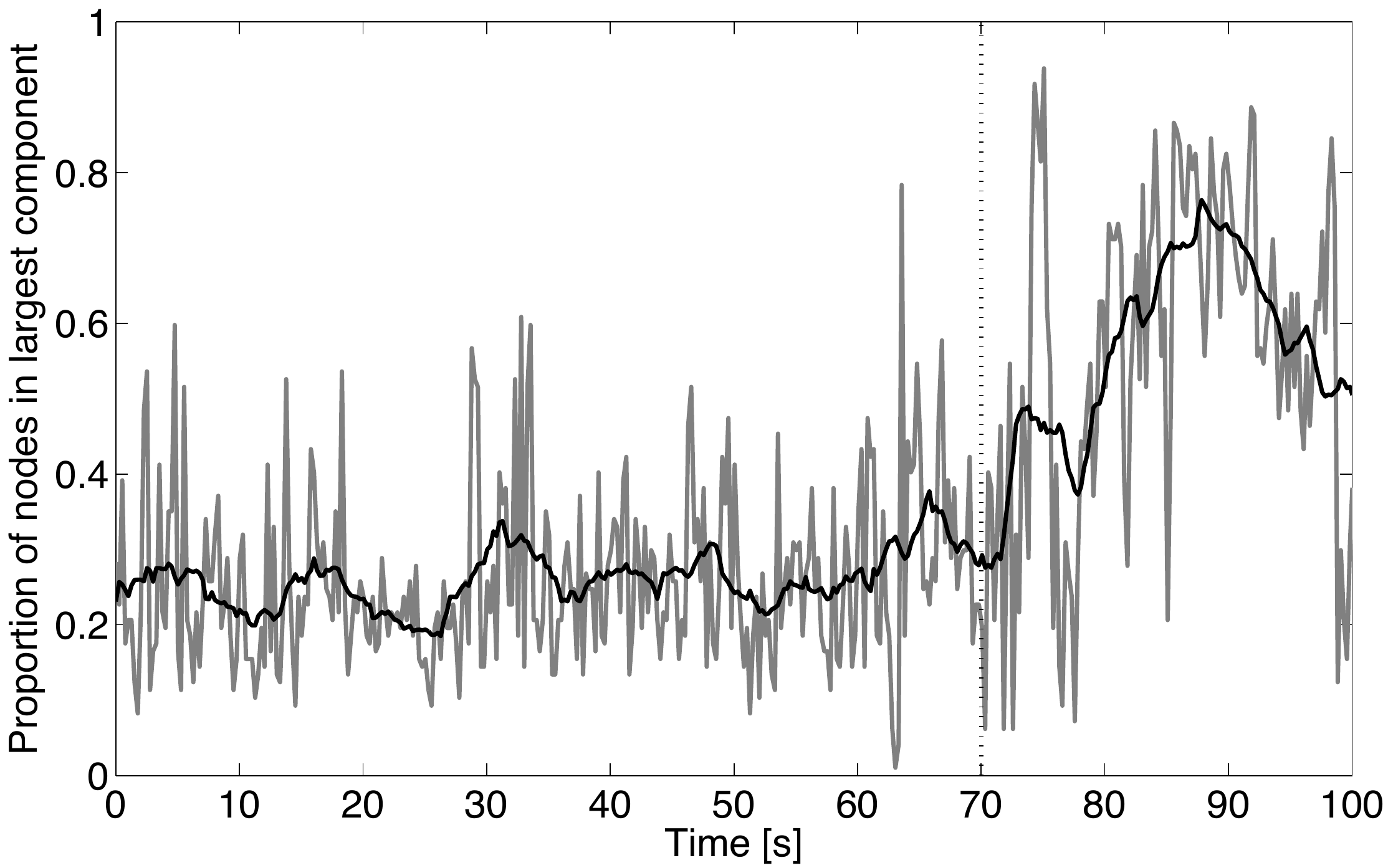}
  \caption{Proportion of nodes in the largest component
    as a function of time for a functional network deduced from the electrocorticogram of
    a single patient with epilepsy during a seizure \citep[see][for
    details]{Kramer2009}. The weighted functional
    networks have been binarized by conducting independent hypothesis tests on
    the maximum absolute values of the cross-correlations over $0.5s$
    windows with 50\% overlap and after correcting for multiple comparisons.
    The black trace is a smoothed version of
    this process. Seizure onset was clinically estimated to occur at
    the vertical dotted line.
    \label{fig:data}}
\end{figure}

\section{Percolation Models}\label{sec:model}
\subsection{Birth/Death Erd\H{o}s-R\'{e}nyi (ER) Process}\label{sec:er}
We first construct a graph-valued stochastic process that exhibits the
Markov property. This provides a realistic model for generating
noisy percolation processes, while maintaining a sufficient level of
computational tractability. We will denote a sequence of graph-valued
random variables on $n$ vertices by
\begin{equation}
    \big\{G_t=(V,E_t):t=0,\ldots,T \big\}.
\end{equation}
At each time step, a single edge is either added or deleted. 
Such a sequence will be said to be \textit{Markov} if its edge sets are
controlled by a Markov chain. We impose this dependence through the
use of a binary random variable, denoted
$\left\{Y_t:t=0,\ldots,T\right\}$, whose state space is
$\{0,1\}$. This Markov chain is characterized by the following
transition probability matrix $P$, for some choices of the birth and
death rates, denoted respectively by $p$ and $q$, and taking values in
$[0,1]$. 
\begin{equation}
\begin{array}{c|c|c}
& Y_{t+1}=0 & Y_{t+1}=1\\
\hline
 Y_t=0 & 1-p & p\\
\hline
 Y_t=1 & q & 1-q
\end{array}
\notag
\end{equation}
Following customary notation, the entries of $P$ will be denoted by
$\p[Y_{t+1}=j|Y_{t}=i]$, with rows summing to one.
The graph-valued Markov chain, $G_{t}$, is then obtained by associating $Y_{t}$
with the addition or deletion of an edge in each edge set, $E_{t}$.
Thus, provided that $p,q\neq0$, it follows that the state space of
this graph-valued Markov chain is the space of
all simple graphs on $n$ vertices, since every graph is reachable
with positive probability. 

In general, $p$ and $q$ are not required to sum to one. It
will be of interest to let $p>q$ in order to study the large-scale
behavior of the $G_{t}$'s as the graph process accumulates edges. 
Moreover, observe that $Y_{t}$ is a (time) \textit{homogeneous} Markov chain, 
since $\p[Y_{t+1}=\ome|Y_{t}=\ome\pri]=\p[Y_{1}=\ome|Y_{0}=\ome\pri]$,
for any $\ome,\ome\pri\in\{0,1\}$ and every $t$.

Now, suppose that there exist $m_{t}:=|E_{t}|$ edges at
time $t$ in $G_t$ and let $X_t(e)$ denotes the `status' of edge $e$ at
time $t$, such that $X_{t}(e)=1$, if that edge is present
and $X_{t}(e)=0$, otherwise. Note that we have
here two different sources of dependence. On one hand, the edges are
dependent on each other, since no more than one edge can be added or
deleted at every time step. On the other hand, the edges are also
dependent over time, since the status of an edge at time $t+1$ depends
on the status of that same edge at time $t$. 

In the sequel, we will concentrate on a special case of this
birth and death process, where we will set $p=1-q$. This leads to simplified
marginal distributions for the edges. Additional details of this birth
and death model are provided in appendix \ref{app:ER details}. 

\subsection{Birth/Death Product Rule (PR) Process}\label{sec:pr}
We extend the standard Achlioptas' framework of PR percolation to 
a birth and death process, by devising death steps. This model is
analogous to the aforementioned ER birth/death model, except for the
choice of the probability distribution of the latent $X_{t}(e)$'s. 
As for the ER birth/death model, a binary random variable, $Y_{t}$, controls
the addition or deletion of edges in each $G_{t}$. However, in the case
of the PR model, the choice of the edge to be added or
to be deleted is not uniform over the $E_{t}$'s. Here, this choice depends
on the modular structure of the graph at time $t$. Therefore, as for
the ER model, we obtain a non-stationary stochastic process. 

\begin{figure}[t]
\centering
\tikzstyle{background rectangle}=[draw=gray!30,fill=gray!10,rounded corners=1ex]
\begin{tikzpicture}[font=\small,scale=.8,show background rectangle]
    \draw[rounded corners=10pt,thick] 
    (-2.75,1.0) -- (-2.75,2.0) -- (-0.25,2.0) -- (-0.25,1.0) -- cycle;
    \draw[rounded corners=10pt,thick] 
    (-2.75,-1.0) -- (-2.75,-2.0) -- (-0.5,-2.0) -- (-0.5,-1.0) -- cycle;
    \draw[rounded corners=10pt,thick] 
    (+2.75,1.0) -- (+2.75,2.0) -- (+1.0,2.0) -- (+1.0,1.0) -- cycle;
    \draw[rounded corners=10pt,thick] 
    (+2.75,-1.0) -- (+2.75,-2.0) -- (+1.10,-2.0) -- (+1.10,-1.0) -- cycle;
    \draw[rounded corners=10pt,thick] 
    (-3.5,+2.5) -- (-3.5,-2.5) -- (+3.5,-2.5) -- (+3.5,+2.5) -- cycle;
    \draw(-1.5,1.5)  node(v11){};
    \draw(-1.5,-1.5) node(v12){};
    \draw(1.5,1.5)   node(v21){};
    \draw(1.5,-1.5)  node(v22){};
    \fill[fill=black](v11) circle (2.0pt); 
    \fill[fill=black](v12) circle (2.0pt); 
    \fill[fill=black](v21) circle (2.0pt); 
    \fill[fill=black](v22) circle (2.0pt); 
    \draw[semithick,dashed] (v11) -- (v12) node[pos=0.5,anchor=west]{$e_{1}$};
    \draw[semithick] (v21) -- (v22) node[pos=0.5,anchor=east]{$e_{2}$};
    \draw(v11) node[anchor=east]{$v_{11}$};
    \draw(v12) node[anchor=east]{$v_{12}$};
    \draw(v21) node[anchor=west]{$v_{21}$};
    \draw(v22) node[anchor=west]{$v_{22}$};
    \draw(+3.5,-2.5)node[anchor=north west]{$G_{t}$};
    \draw(-2.5,1.0)node[anchor=north east]{$C_{11}$};
    \draw(-2.5,-1.0)node[anchor=south east]{$C_{12}$};
    \draw(+2.5,1.0)node[anchor=north west]{$C_{21}$};
    \draw(+2.5,-1.0)node[anchor=south west]{$C_{22}$};
    \draw[gray!10] (-4.5,-3.0) -- (+4.5,-3.0);
    \draw[gray!10] (-4.5,+3.0) -- (+4.5,+3.0);
\end{tikzpicture}  
\caption{Birth/death steps for the product rule (PR) in the Achlioptas'
      model of percolation. In this example, edge $e_{2}$ was born \textit{before}
      edge $e_{1}$, since $|C_{11}||C_{12}| > |C_{21}||C_{22}|$. 
      Therefore, when $e_{1}$ and $e_{2}$ are selected during a death step, 
      $e_{2}$ is discarded \textit{after} $e_{1}$. This death step
      specification ensures that births and deaths constitute genuine reverse
      PR operations.
      \label{fig:ordering}}
\end{figure}
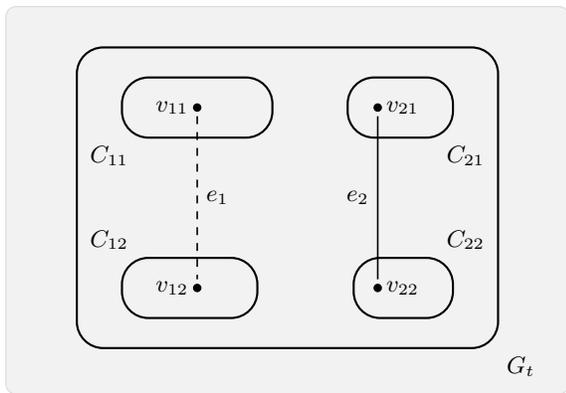

Assuming that $Y_t=1$, the addition of a new edge is
conducted by uniformly choosing two candidate vertex pairs among all
the edges in $E^{C}_{t}$, the complement of the edge set,
$E_{t}$. These two candidate edge pairs are denoted by
$e_1:=(v_{11},v_{12})$ and
$e_2:=(v_{21},v_{22})$, and satisfy $X_{t}(e_{1})=0$ and
$X_{t}(e_{2})=0$, since $e_{1},e_{2}\in E^{C}_{t}$, as in Figure
\ref{fig:ordering}. We then evaluate
the size of the connected components to which $v_{11}$, $v_{12}$, $v_{21}$ and
$v_{22}$ belong. These four connected
components are denoted by $C_{11}$, $C_{12}$, $C_{21}$, and
$C_{22}$, respectively. Then,
following \citet{Achlioptas2009}, we
apply the following product rule: If $|C_{11}||C_{12}|<|C_{21}||C_{22}|$, 
then $X_{t+1}(e_1) = 1$; otherwise, $X_{t+1}(e_2) = 1$.

Conversely, the death or deletion of an edge is handled in a symmetric manner. 
When $Y_t=0$, we uniformly select two candidate edges
from $E_{t}$. These vertex pairs are denoted 
$e_1:=(v_{11},v_{12})$ and $e_2:=(v_{21},v_{22})$
and satisfy $X_t(e_1)=1$ and $X_t(e_2)=1$, since $e_{1},e_{2}\in E_{t}$.
Next, we set $X_t(e_1)=0$ and $X_t(e_2)=0$, in
order to compute the size of the connected components to which
$v_{11}$, $v_{12}$, $v_{21}$ and
$v_{22}$ would belong to, if these edges were absent. This is done in
order to ensure that the deletion of an edge exactly corresponds to the
reverse operation of the addition of an edge under PR.
Now, after having deleted these edges and computed the sizes of
$C_{11}$, $C_{12}$, $C_{21}$, and $C_{22}$; we decide which edge should
re-enter $G_t$, in order to produce $G_{t+1}$. Such a decision is also
based on the PR, such that if $|C_{11}||C_{12}|<|C_{21}||C_{22}|$, 
then $X_{t+1}(e_2) = 0$; otherwise, $X_{t+1}(e_1) = 0$.
\begin{figure}[t]
\centering
\tikzstyle{background rectangle}=[draw=gray!30,fill=gray!10,rounded corners=1ex]
\begin{tikzpicture}[font=\small,scale=1.0,show background rectangle]
    \draw (-2.0,0)node[draw,minimum size=1.0cm](x1){$G\as_{t-1}$};
    \draw (0,0) node[draw,minimum size=1.0cm](x2){$G\as_{t}$};
    \draw (2.0,0) node[draw,minimum size=1.0cm](x3){$G\as_{t+1}$};

    \draw (-2.0,2.0)node[draw,circle,minimum size=1.0cm](l1){$G_{t-1}$};
    \draw (0,2.0) node[draw,circle,minimum size=1.0cm](l2){$G_{t}$};
    \draw (2.0,2.0) node[draw,circle,minimum size=1.0cm](l3){$G_{t+1}$};

    \draw[thick,->,dashed] (-3.5,2.0) -- (l1);
    \draw[thick,->] (l1) -- (l2);
    \draw[thick,->] (l2) -- (l3);
    \draw[thick,->] (l1) -- (x1);
    \draw[thick,->] (l2) -- (x2);
    \draw[thick,->] (l3) -- (x3);
    \draw[thick,->,dashed] (l3) -- (+3.5,2.0);

    \draw[color=gray!10](-3.2,-.8)--(3.2,-.8);
    \draw[color=gray!10](-3.2,+2.8)--(3.2,+2.8);
\end{tikzpicture}
\caption{Directed Acyclic Graph (DAG) representation of the hidden
      Markov process combining a \textit{latent} stochastic graph
      process in the first row denoted by $G_{t}$, with an \textit{observed}
      stochastic graph process contaminated by noise in the second
      row, denoted by $G\as_{t}$. Directed arrows indicate
      probabilistic dependence, such that the distribution of the
      observed $G_{t}\as$ depends on the value taken by the latent
      graph, $G_{t}$.
      \label{fig:latent}}
\end{figure}
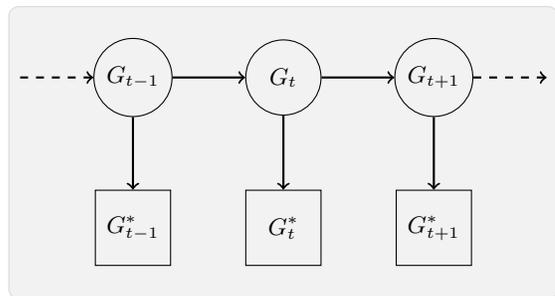

This choice of specification for the death step
ensures that the \textit{ordering} of the creation and deletion of
edges are symmetrical. Given a sequence of two edges
$\{e_{1},e_{2}\}$ successively born during two time steps of $G_{t}$, if we
encounter a death step, where both $e_{1}$ and $e_{2}$ are selected, we
would then delete these edges in the reverse order, by eliminating $e_{2}$
before $e_{1}$. This order-preserving property is illustrated in
Figure \ref{fig:ordering}. This constraint ensures that births and
deaths are genuine reverse PR operations. In addition, observe that, as for the ER
percolation process, this chain is \textit{irreducible}, in the sense
that there is positive probability of transitioning from any given edge
configuration to any other in the space of the edge sets of $G$.

\subsection{Hidden Markov Graph Model}\label{sec:hmm}
Next, we assume that there exists a time-independent error process, which
produces at each time point an \textit{observed} edge status
$X_{t}\as(e)$. This stochastic process is governed by 
two additional parameters $\alpha$ and $\beta$, whose behavior 
can be described using a traditional `confusion matrix', such
that for any $\alpha,\beta\in[0,1]$, we have
\begin{equation}
\begin{array}{c|c|c}
 & X_t^*(e)=0 & X_t^*(e)=1\\
\hline
X_t(e)=0 & 1-\alpha & \alpha \\
\hline
X_t(e)=1 & \beta & 1-\beta \\
\end{array}
\notag
\end{equation}
\begin{figure}[t]
  \centering
  \includegraphics[width=8.5cm]{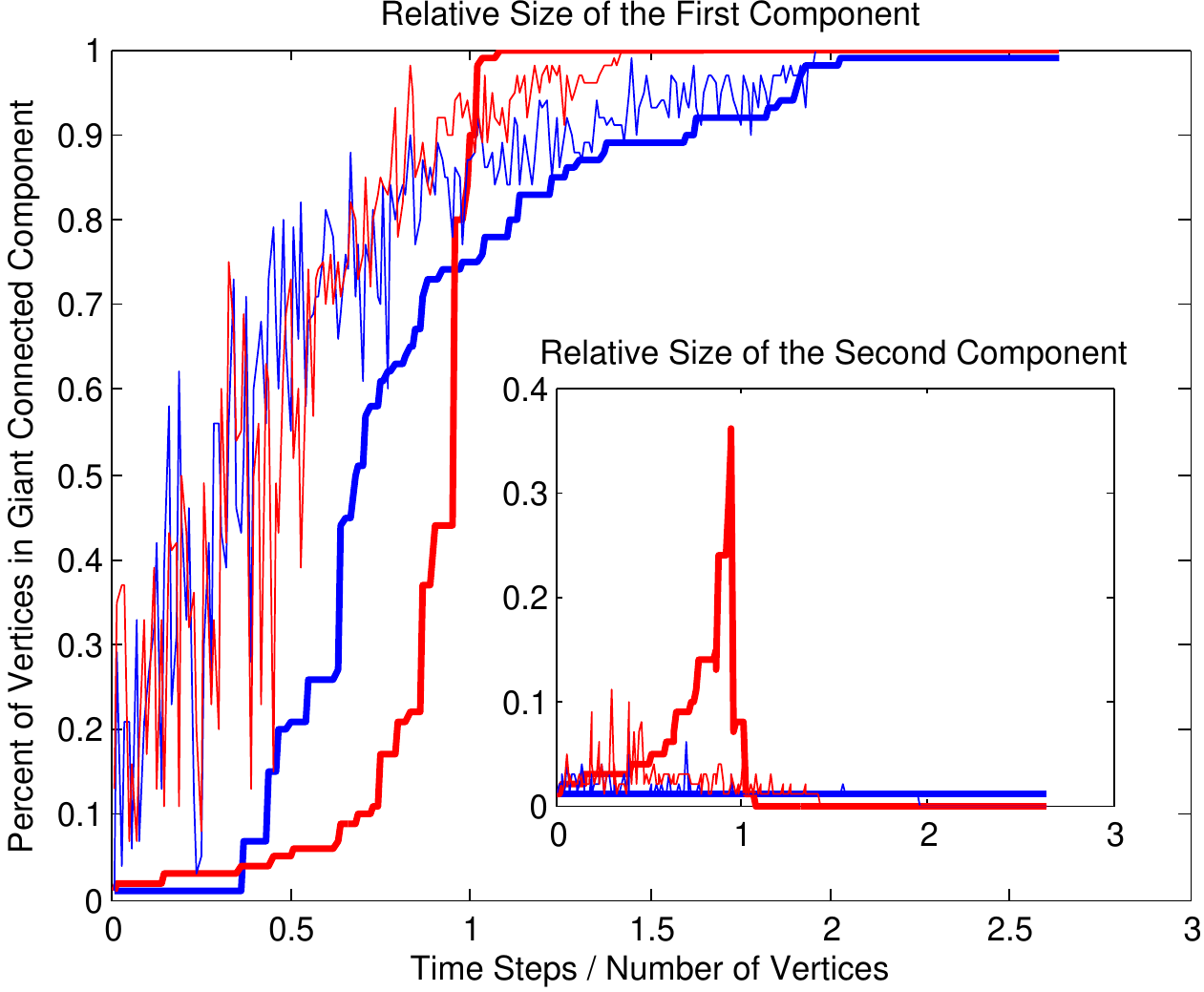}
  \caption{Percentage of vertices in the giant connected component
    (GCC) in the main window, and the corresponding sizes
    of the second largest component in inset, for a birth rate of
    $p=1$ and death rate of $q=0$, for the \textbf{ER (blue)} and
    \textbf{PR (red)} percolation models. In bold, these results are
    reported for a noise-free
    model, whereas the thin lines represent noisy simulations with 
    Type I/II error rates of $\al=.0125$, $\be=.01$. Observe that the
    two models become almost indistinguishable when noise is added to
    these processes. 
    \label{fig:criteria}}
\end{figure}

The $X_{t}(e)$'s and $X\as_{t}(e)$'s are here treated as \textit{latent}
and \textit{observed} stochastic processes, respectively, and $\alpha$ and
$\beta$ can therefore be interpreted as the Type I (false positive)
and Type II (false negative) error probabilities. Combining the
graph-valued Markov latent process with
this time-independent error process, we obtain a
graph-valued \textit{hidden Markov process}, as described in figure
\ref{fig:latent}. From this schematic representation, one can
immediately see that the observed graphs denoted $G\as_{t-1}$, $G\as_{t}$ and
$G\as_{t+1}$ are conditionally independent, given the latent
graph process, $G_{t}$. 

For the ER model, under the assumption that $p=1-q$,
these two stochastic graph processes can be combined by taking into
account the time-dependence of the
$X\as_{t}(e)$'s. In this case, the corresponding transition matrix linking the
observed and latent processes is available in closed-form. Details of
these derivations are provided in appendix \ref{app:ER details}.
\begin{figure}[t]
  \centering
  \includegraphics[width=8.5cm]{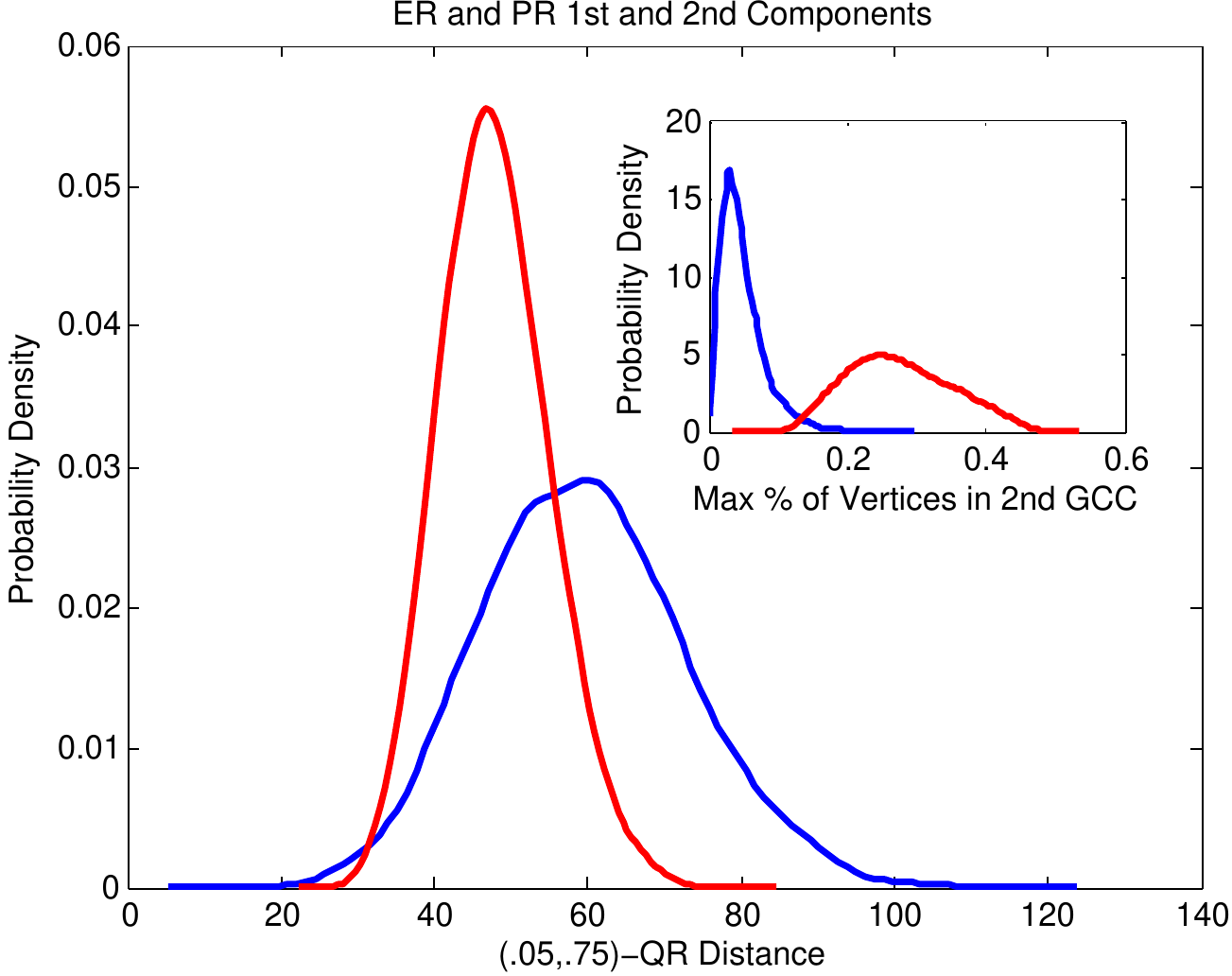}
  \caption{Densities of the QD$(.05,.75)$ 
    of the first component's size in the main window, and 
    densities of the maximal size of the second component (inset) are reported 
    for the \textbf{ER (blue)} and \textbf{PR (red)} percolation
    models, based on 1,000 draws from the distributions of these two
    (noise-free) models. Note that the difference in scales of the
    density values of the $y$-axes in these two figures is due to the
    difference in scales of the $x$-axes.
    \label{fig:densities}}
\end{figure}

\section{Detecting Explosive Percolation}
Explosive percolation is expected to produce a sharper
phase transition than a typical ER percolation. When considering noisy
observations, however, detecting such differences through visual
inspection only is hard. Figure \ref{fig:criteria} illustrates this problem, by
comparing noisy and noise-free graph sequences for both explosive and
ER models. Beyond visual inspection, the problem of discriminating between these two models of
percolation can be formulated as a hypothesis-testing problem:
The null hypothesis, denoted $H_{0}$, states that the observed
process corresponds to an ER percolation, whereas the alternative
hypothesis, $H_{1}$, is that the observed process does not correspond
to this type of percolation model.

To proceed with this hypothesis-testing problem in practice, we specify
a population parameter summarizing the percolation process, say $\theta^{\op{ER}}$ and
$\theta$, for the ER and target models, respectively. This leads
to a hypothesis test of the form,
\begin{equation}
  H_{0}: \theta^{\op{ER}} = \theta, \qquad\te{and}\qquad
  H_{1}: \theta^{\op{ER}} \neq \theta.
  \notag
\end{equation}
Several population parameters could be used for the purpose of
discriminating between these two models of percolation. A natural
candidate for such parameters would be a measure of the sharpness of
the transition of the first component's size. 
The main panel of Figure \ref{fig:criteria}, however,
suggests that this population parameter will not have sufficient
discriminatory power, when confronted with a substantial amount of
observational noise. 

Therefore, as a second candidate population parameter, we consider the following natural
extension: The size of the \textit{second largest} component. This appears to
provide a more sensitive marker of the sharp phase transition exhibited
by explosive percolation models. This `second-order' property
is motivated by the mechanism of the product rule, which underlies the
Achlioptas' graph process, which we will use as representative of the
alternative hypothesis, $H_{0}$ (i.e., reducing our testing problem
to the comparison of two point null hypotheses).
In what follows, we will show that the use of the size of the second
largest component as a statistical marker to distinguish the two percolation
regimes exhibits greater discriminatory power, than a statistic solely
based on the first component. 

The differences between the candidate percolation models are
therefore quantified using two criteria: (i) a quantile difference
(QD) of the distribution of the size of the GCC, and (ii) the maximal size
of the second component over the entire time period. 
These two criteria are formally defined as follows.
Given the graph process, $G_{t}=(V,E_{t})$, and denoting the
vertex subset of the largest component in $G_{t}$ by $S_{1,t}$, we
define the cumulative edge function as the \textit{cardinality} of
$S_{1,t}$, normalized by the maximal number of edges in the graph,
such that 
\begin{equation}
   F(t):= \frac{|S_{1,t}|}{\binom{n}{2}}.
   \notag
\end{equation}
Although this function is not a cumulative distribution
function (CDF), one can nonetheless uniquely define quantiles using the
standard definition of quantiles for the CDFs of discrete random variables;
such that for any $x\in[0,1]$, we have
\begin{equation}
  Q(x) := \min_{t=1,\ldots,T}\lt\lb t: F(t)\geq x\rt\rb,
  \notag
\end{equation}
where $T$ is the maximal number of time steps in the graph process. 
In this paper, we are especially interested in quantile differences of
the following form, which can be treated as a generalization of the
classical interquartile range, 
\begin{equation}
  \op{QD}(x_{1},x_{2}) := Q(x_{2}) - Q(x_{1}). 
  \notag
\end{equation}
One can observe from Figure \ref{fig:data} that the size of the GCC,
in the functional networks associated with a seizure,
rarely exceeds 80$\%$. Therefore, we have here
adopted a quantile difference that reflects the range of the
distribution of the GCC in this practical setting. Thus, the criterion
of interest will be the following, $\theta_{\op{QD}}:=Q(.75)-Q(.05)$. 
This parameter quantifies the steepness of the phase transition, 
the larger the QD criterion, the longer the
transition to a fully connected graph. 

As a second criteria to distinguish the two percolation regimes, we
consider the maximal size attained by the second largest
component over the entire time period of the dynamic network observation.
If one defines the vertex set of the second largest component at time
$t$ by $S_{2,t}$, this second criterion can be expressed as
$\theta_{\op{Sec}}=\max\lt\lb |S_{2,t}|:t=1,\ldots,T\rt\rb$. This
quantity is expected to constitute a good
marker of the steepness of the phase transition,
since it reflects the extent of separation of the graph process into
large connected subgraphs. Indeed, a direct consequence of the
Achlioptas' construction rule is that by inhibiting the growth of a single
large component, we necessarily increase the production of several subcomponents.

Statistical inference on these two criteria is then drawn using a
Monte Carlo hypothesis test. Letting the parameter $\theta:=\theta_{\op{QD}}$, and
selecting the candidate percolation to be drawn from a PR process, we
consider the following null and alternative hypotheses,
\begin{equation}
  H_{0}: \theta^{\op{ER}} = \theta^{\te{PR}}, \qquad\te{and}\qquad
  H_{1}: \theta^{\op{ER}} > \theta^{\te{PR}},
  \notag
\end{equation}
respectively. The direction of this test is justified by the fact that
we expect explosive percolation to occur rapidly, and thus to
exhibit a smaller amount of variability in the size of its GCC, when transitioning to a
fully connected graph. Our second criterion, by contrast, is tested in
the opposite direction, since we naturally 
anticipate the PR process to be characterized by 
a \textit{larger} maximal second component. Thus, for
$\theta:=\theta_{\op{Sec}}$, the alternative hypothesis becomes
$H_{1}: \theta^{\op{ER}} < \theta^{\te{PR}}$.
\begin{figure}[t]
  \begin{flushleft}\textbf{(a)}\end{flushleft}\vspace{-.5cm}  
  \includegraphics[width=8.5cm]{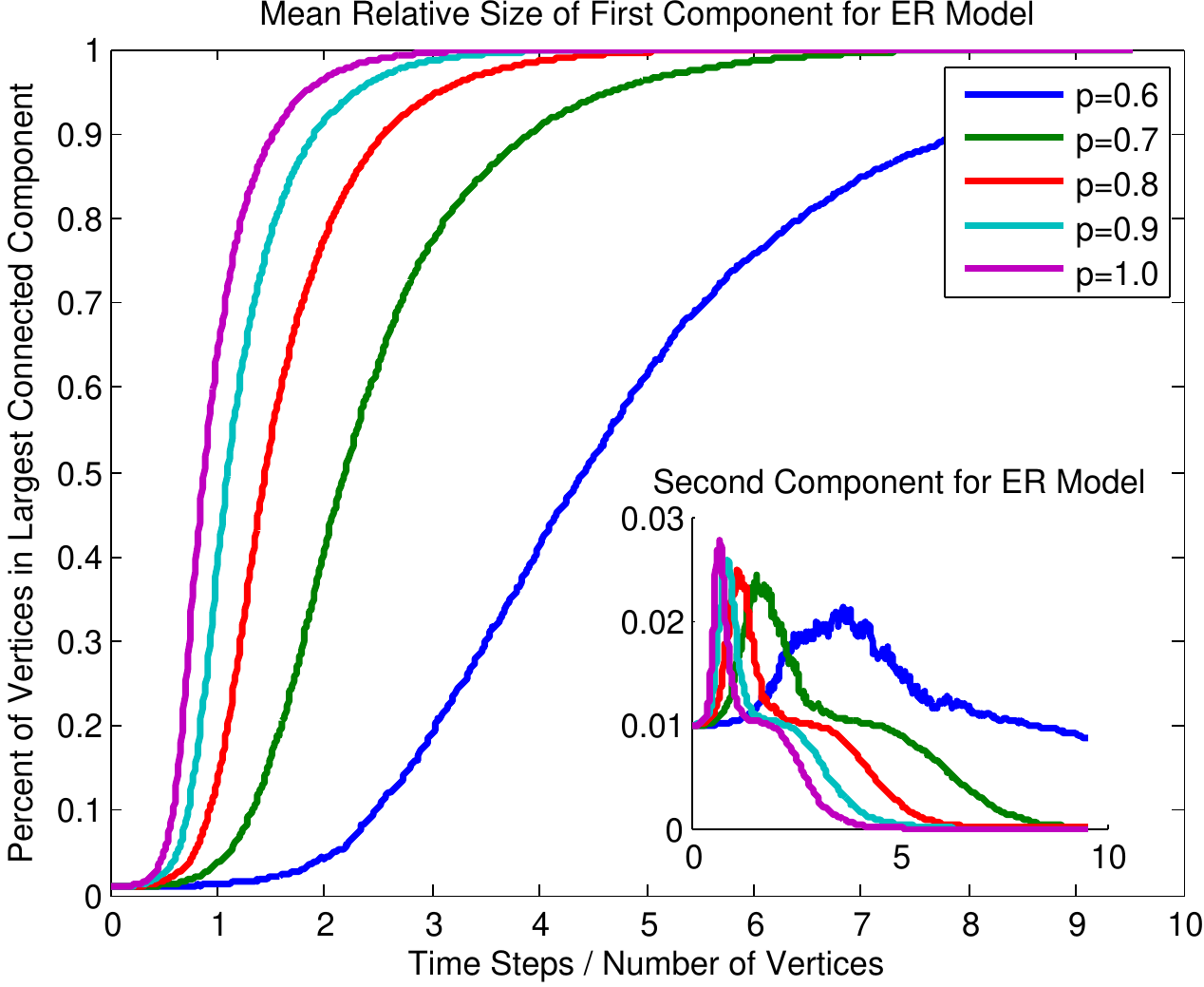}\\
  \begin{flushleft}\textbf{(b)}\end{flushleft}\vspace{-.5cm}  
  \includegraphics[width=8.5cm]{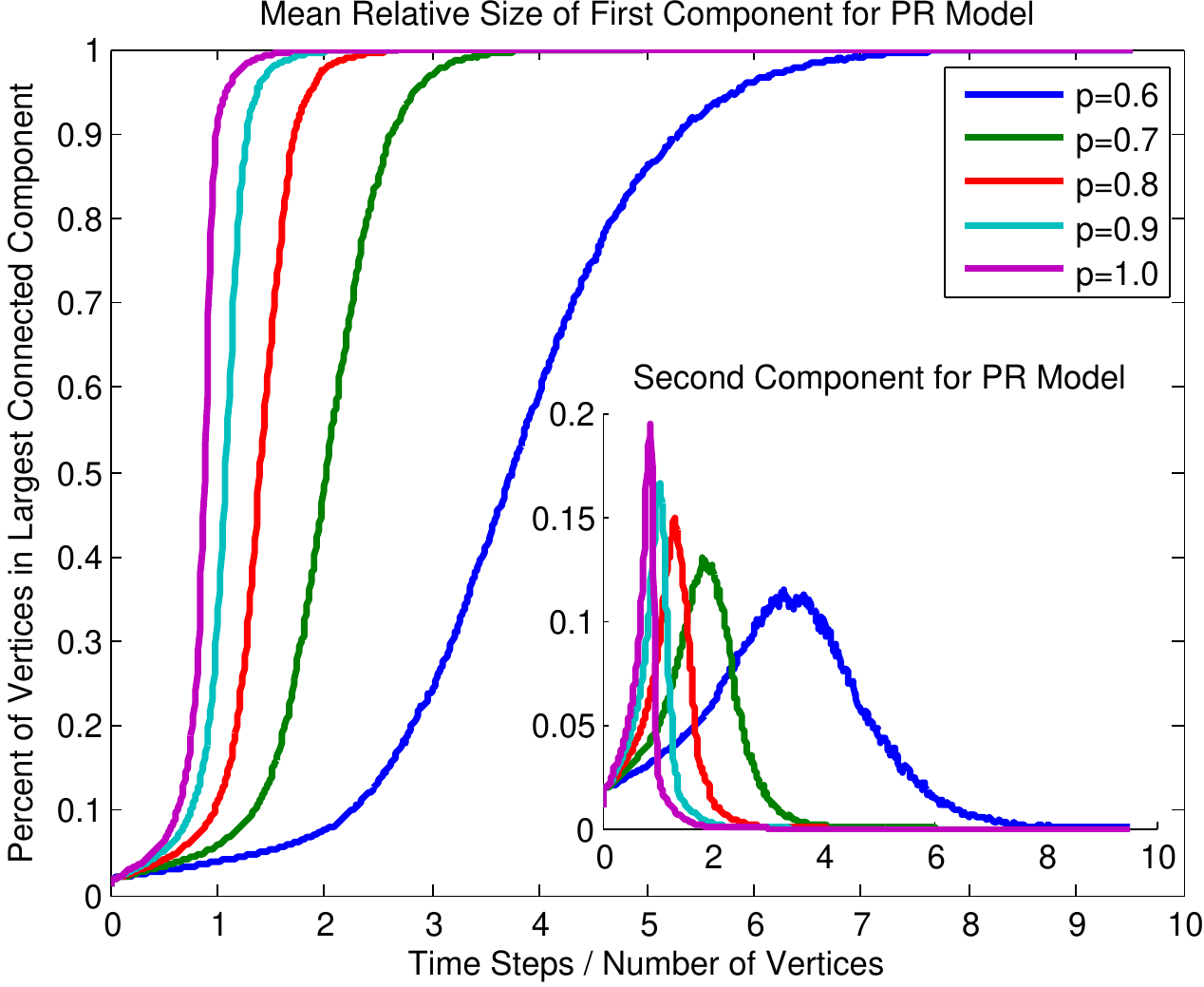}
  \caption{Percentage of vertices in giant connected component (GCC)
    with respect to time, under the ER and PR models, in panels (a) and
    (b), respectively; for different choices of birth rates, $p$. 
    In each panel, the corresponding sizes of the second largest
    component are also reported in inset. The edge death rate is here
    dependent on the birth rate, such that $q=1-p$. Each curve
    represents the mean of 1,000 different simulations. 
    \label{fig:meanCurves}}
\end{figure}

In the results reported in this paper, the distributions of the ER and PR graph
processes are known. It therefore suffices to simulate from these
densities in order to construct the distribution of the two test
statistics at hand. This procedure is illustrated in figure
\ref{fig:densities}. We are especially interested in the 
discriminatory powers of these statistics, and we will therefore
compare their respective merits, using the true positive and
false positive rates, within a receiver operating characteristic (ROC)
framework. For presentational convenience, the distributions of
interest were smoothed using a normal density kernel, before computing
the ROC curves and corresponding \textit{areas under the curves} (AUCs). The
computation of the AUCs allows us to summarize the
differences between the models over the entire time period. 

\section{Results}\label{sec:result}
\subsection{Birth/Death Processes}
\begin{figure}[t]
  \centering
  \includegraphics[width=8.5cm]{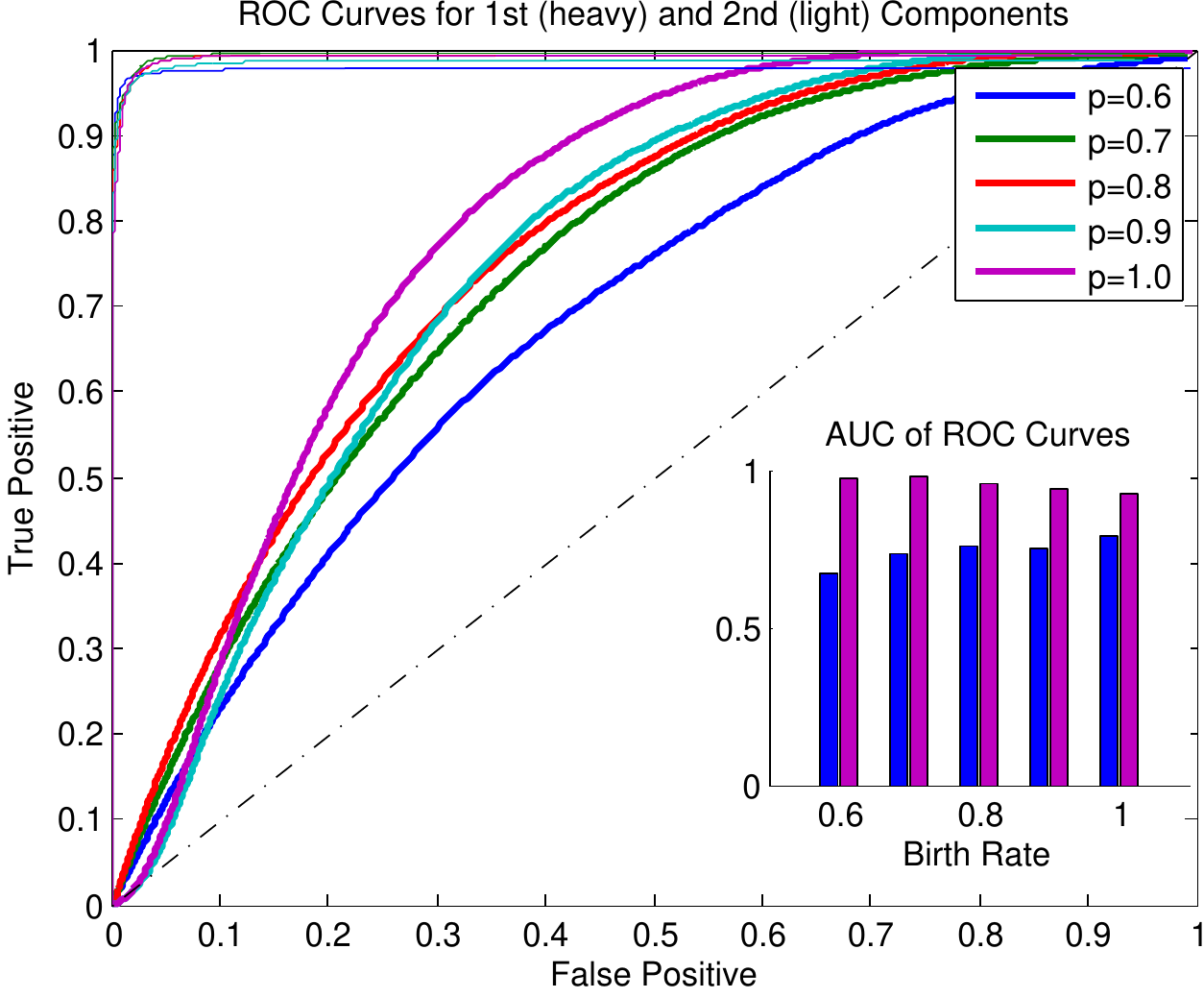}
  \caption{Receiver operating characteristic (ROC) curves based on the QD
    of the first component's size (bold lines) and on the maximal size of the
    second component (thin lines). Each curve was produced using 1,000
    simulations from each percolation model. In inset, the areas under the
    curves (AUCs) of these different ROC curves are compared for
    different choices of the birth rate, $p$, with the blue and purple bars
    denoting the ER and PR models, respectively. 
    \label{fig:auc}}
\end{figure}
The ER and PR percolation models were simulated on graphs of $n=100$
vertices. We first explored the effect of varying the birth and death rates on
the behavior of the two statistical criteria of interest, under the ER and PR
models. The results of these Monte Carlo simulations are reported in
Figure \ref{fig:meanCurves}, where each curve is the mean 
of 1,000 different synthetic data sets. In these simulations, the death
rate was set to $q=1-p$, and therefore the value of $p$ controls both
the birth and death rates. 

The main effect of a change in $p$ is to delay percolation, and to
diminish the steepness of the phase transition. Observe that as $p$
decreases, percolation tends to occur at a later time step in both the
ER and PR models (Figure \ref{fig:meanCurves}). In particular, for
the lowest birth rate  that
we investigated $(p=0.6)$, the ER model did not produce a fully connected graph
within the number of iterations considered, as can be
seen from Figure \ref{fig:meanCurves}(a). When $p$ was set to values
equal to or less than $0.5$, no phase transition could be observed, and these
results are not reported. 

The size of the second component was similarly affected by changes in
$p$. Decreasing the birth rate delayed the time at which the
size of the second component attained its highest value. Moreover,
lower values of $p$ also yielded second components with smaller
maximal sizes, under both the ER and PR models. Interestingly, 
we note that the time points at which the second components 
reach a maximal size tend to coincide in both models. Thus, it would
be difficult to distinguish between these two percolation models on
the sole basis of the \textit{timing} of the occurrence of the maximal size
of the second components. By contrast, the relative maximal \textit{size}
of the second components in the ER and PR models differ by
approximately one order
of magnitude, thereby providing a natural criterion for discriminating
between these two types of percolation, as can be observed by
comparing the inset figures in \ref{fig:meanCurves}(a) and 
\ref{fig:meanCurves}(b).

We formally quantified these differences in discriminatory powers 
by studying the ROC curves of these two criteria under different choices of
$p$ (Figure \ref{fig:auc}). The maximal size of the second component substantially
outperforms the relative size of the first component, for all values of $p$. The stark
difference between these discriminatory criteria can be understood by
considering the amount of overlap of the distributions of these two
criteria in Figure \ref{fig:densities}. Whereas the distributions of
the QD of the size of the GCC under the two models exhibit a large
amount of overlap; the distributions of the maximal size of the second
component, by contrast, share very little common support. These
differences in support account for the substantive gains in
discriminatory power by the maximal size of the second component, 
reported in Figure \ref{fig:auc}.
 
In addition, we note the QD of the size of the first component was
more sensitive to choices of $p$ than the maximal size of the second component. 
As $p$ diminishes, it becomes increasingly more difficult to
discriminate between the ER and PR percolation models, using the QD
of the size of the first component. This suggests
that this criterion is more sensitive to a non-zero
death rate, than the maximal size of the second component, which
provides further support for the use of this latter criterion, in
practice. 
\begin{figure}[t]
  \begin{flushleft}\textbf{(a)}\end{flushleft}\vspace{-.5cm}  
  \includegraphics[width=8.5cm]{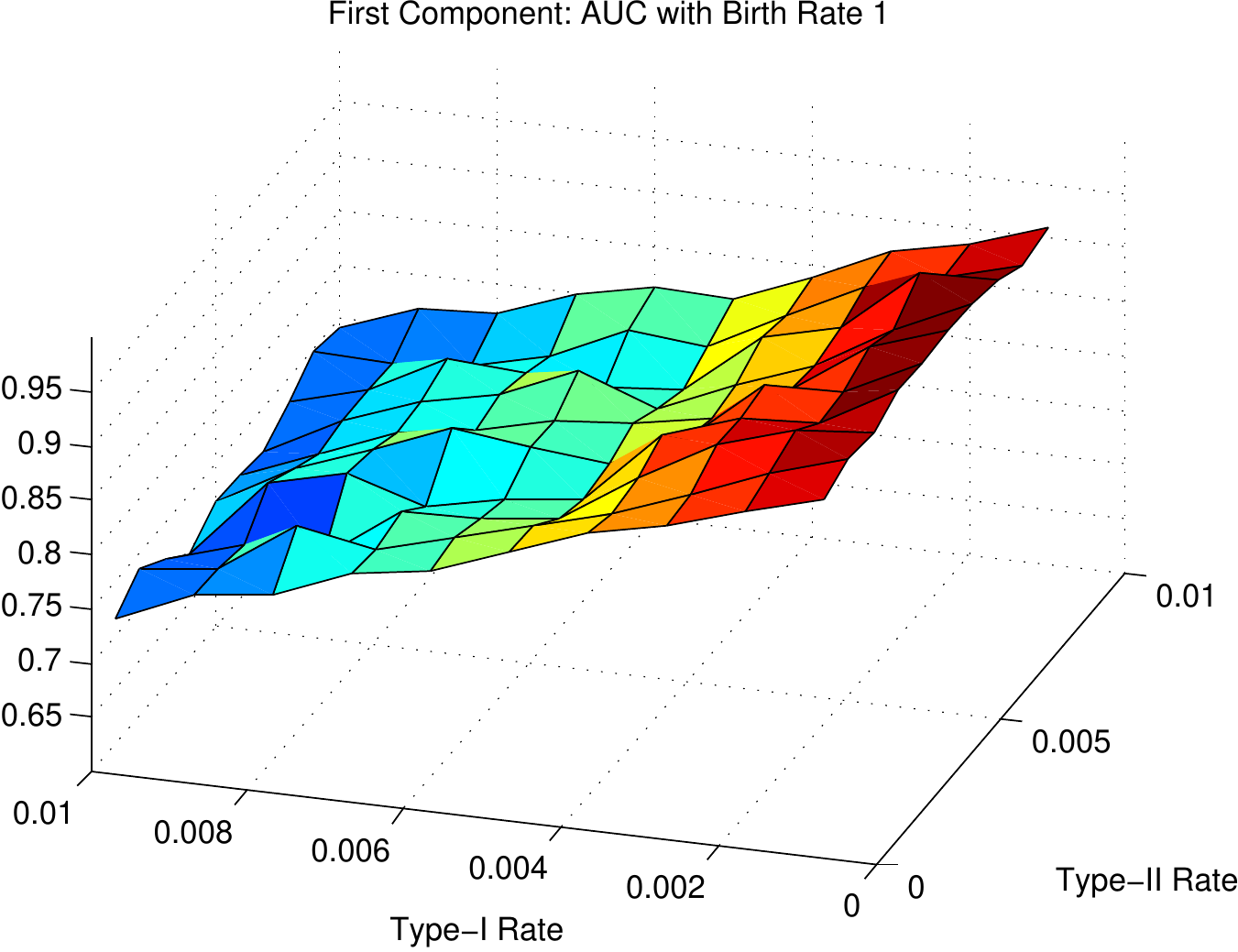}
  \begin{flushleft}\textbf{(b)}\end{flushleft}\vspace{-.5cm}  
  \includegraphics[width=8.5cm]{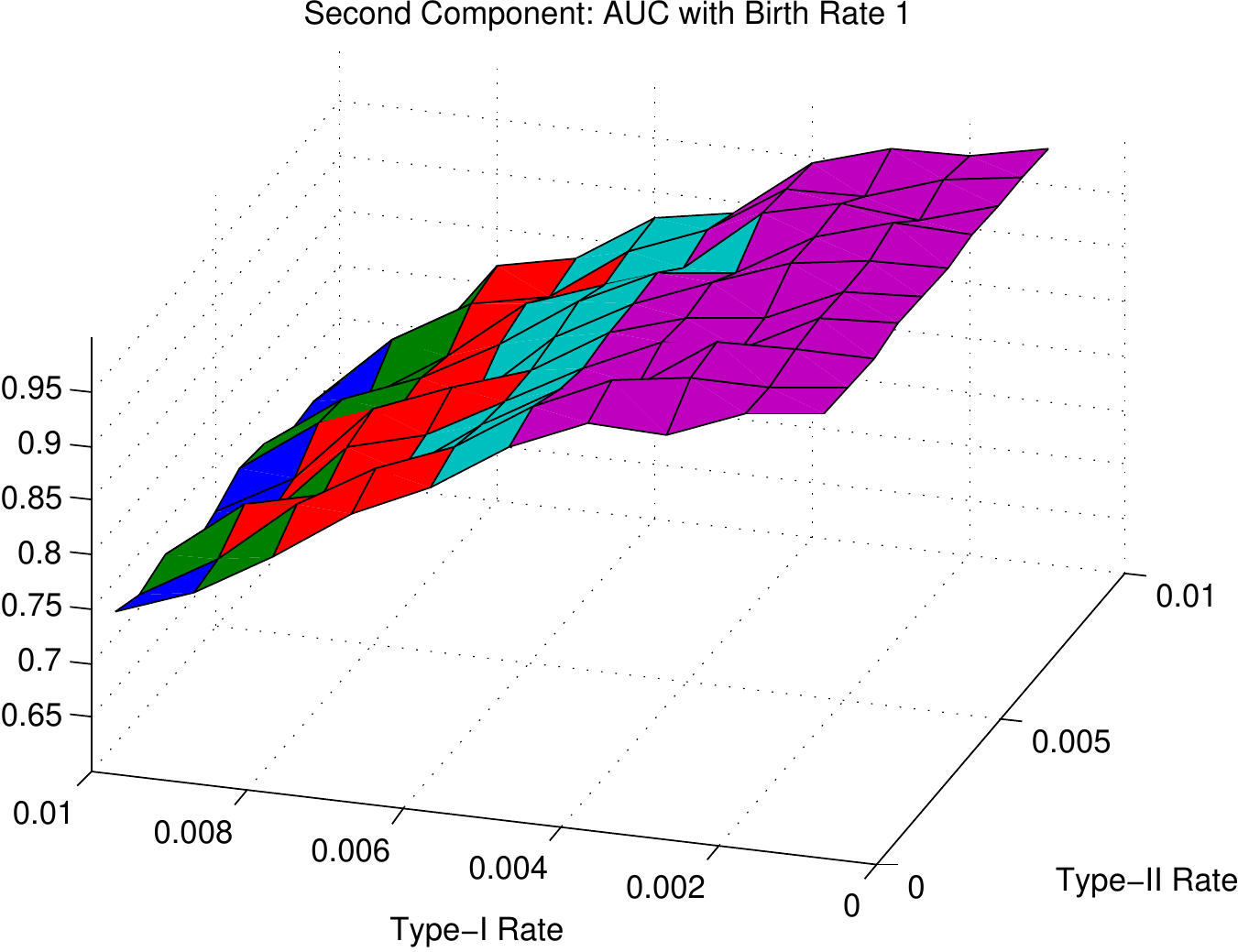}
  \caption{Areas under the curve (AUCs) for ROC curves for $10^5$
    model simulations based on the QD
    of the first component's size in panel (a), and based on the
    maximal size of the second component in panel (b). We have 
    varied the Type I and Type II error rates within a hidden Markov
    graph model framework. The birth/death rates have been kept
    constant with $p=1$. 
    \label{fig:noisy}}
\end{figure}
\subsection{Percolation under Noise}
Secondly, we considered the effect of introducing noise in these
models. The results reported in Figure \ref{fig:noisy} were produced
using our proposed hidden Markov graph model, and are averaged over 1,000
simulations. We were especially interested in the effect of Type I
and Type II errors on our ability to discriminate between classical
and explosive percolation, using the two criteria under scrutiny. 
Both the Type I and Type II error rates were made to vary between
$0$ and $.01$. 

From Figure \ref{fig:noisy}, one can observe that the two types of
errors had markedly different effects on the AUCs of the two
discriminatory criteria. Introducing Type I errors led to a substantial
diminution of the AUCs for both the QD of the size of the first
component in (a), and the maximal size of the second component in
(b). In particular, note that the two criteria reached equivalent
levels of discriminatory power for $\al=0.01$. Thus, although the
maximal size of the second component remains a more useful criterion
for distinguishing between the ER and PR models than the QD of the
size of the first component, these two criteria exhibit comparable
performance, under a moderate amount of Type I error.

The impact of increasing the Type II error rate on the behavior of
these two criteria was negligible. Introducing false negatives in
the ER and PR models slightly increased the AUCs of both the QD of
the first component's size, and the maximal size of the second
component. Thus, large Type II error rates may be 
marginally advantageous for discriminating between these two models of
percolation, under the scenarios studied. 

\section{Discussion and Conclusions}\label{sec:conclusion}
In this paper, we have extended existing models of percolation, by
allowing for edge deletion steps and noisy observations. These
modeling extensions have been articulated within a hidden Markov graph
process, which builds links with the existing literature on the
statistical properties of this family of models
\citep{Flocchini2012,Casteigts2012,Lentz2013}.
Moreover, we have compared different summary statistics for
distinguishing between the ER and PR percolation models. Overall, for
different birth and death rates, and for a range of noise levels, the
maximal size of the second component was found to have greater
discriminatory power than the QD of the size of the GCC. 

Several methodological challenges remain before such models can be
directly used for percolation detection on real-world data. 
Throughout this paper, we have considered the QD of the size of the
first component, using a particular choice of quantiles for this
discriminatory criterion. 
In practice, an optimal choice of quantiles for quantifying the
steepness of such phase transitions may be motivated by different
factors, including (i) the range of the observations, and (ii) the
need for early detection. We discuss these two practical aspects,
in turn. 

Firstly, note that when considering real-world applications, we rarely observe fully
connected networks. In the data reported in Figure \ref{fig:data}, for
instance, the size of the GCC encompasses at most 90\% of the edges in
the saturated network. The choice of the quantile interval
of interest for the first component will be therefore automatically
constrained by the range of the observations in the data at hand. 
Therefore, as in sequential detection analysis, the statistical objective is to
detect the outcome, on the basis of as little data as possible. Such
constraints would naturally lead to a relatively narrow quantile range. 

Secondly, in the context of clinical neuroscience and with
particular emphasis on the prevention of a seizure; the detection of
a percolation regime may be linked with patients' health and survival.
In such cases, early detection will usually be
favored, as this is likely to be associated with desirable
clinical outcomes. Explosive percolation, such
as the Achlioptas' PR process studied in this paper, 
is consistent with the sudden manifestation of a seizure as a highly
synchronized event. Classifying models of
percolation may then be utilized to deepen understanding of seizures
in epilepsy, and a statistical identification of explosive phase transitions
may facilitate immediate, targeted intervention (e.g., an electrical
stimulus).

Further work in this area could be focused on estimating a percolation
model from a given sequence of observed networks. In this sense, this
work also contributes to the growing literature on time-indexed graph
processes \citep{Grindrod2009}. In such cases, the
birth and death rates will need to be estimated, as well as the Type I
and Type II error probabilities. These different parameters may not be
fully identifiable from the data, and further constraints are likely
to be necessary, in order to discriminate between the two
percolation models considered in this paper. Such estimation, however,
may be amenable to a Bayesian formulation, as commonly implemented
for hidden Markov models \citep{Ephraim2002}.

\appendix
\section{Details of Birth/Death ER Process}\label{app:ER details}
Here, we describe the closed-form formulas of the
probabilities of edge inclusion and edge deletion in the observed graph
processes under the ER model. These analytic results are
obtained by assuming that the birth and death probabilities are
straightforwardly related, such that $p=1-q$. Such derivations may be
useful for other authors, who may want to replicate these results, or
extend the applications of the noisy model of percolation. 

In this birth and death graph process, each edge is treated separately by
integrating out the dependence
of all other edges in the graph, and considering the marginal
distribution of every $X_{t}(e)$. As before, we will here refer to
$X_{t}(e)$ as the \textit{latent} edge status, and $m_{t}:=|E_{t}|$
will indicate the number of edges in the graph at time $t$. Given the
Markov random variable $Y_{t}$, the conditional transition matrix for
every $X_{t+1}(e)$, given some value of $Y_{t}$ takes the following form,
\begin{equation}
\begin{array}{c|c|c}
& X_{t+1}(e)=0 & X_{t+1}(e)=1\\
\hline
X_t(e)=0 &
\frac{\binom{n}{2}-m_{t}-I\left\{Y_t=1\right\}}{\binom{n}{2}-m_{t}} &
\frac{I\left\{Y_t=1\right\}}{\binom{n}{2}-m_{t}} \\
\hline
X_t(e)=1 & \frac{I\left\{Y_t=0\right\}}{m_{t}} &
\frac{m_{t}-I\left\{Y_t=0\right\}}{m_{t}} \\
\end{array}
\notag
\end{equation}
where $\binom{n}{2}$ denotes the number of edges in a saturated
graph of size $n$, and where $I\lb f(x)\rb$ is the
indicator function, which takes a value of $1$ if $f(x)$ is true and
0, otherwise. 

In this paper, we have concentrated on a special case of this
birth/death process, where we have set $p=1-q$. This choice of $p$ and $q$
leads to the following characterization of the $Y_{t}$ process,
\begin{equation}
  \begin{aligned}
   \p\lt[Y_{t+1}=1\rt] &= \p\left[Y_{t+1}=1|Y_t=0\right] \\
   &= \p\left[Y_{t+1}=1|Y_t=1\right] = p,
 \end{aligned}
 \label{eq:independence}
\end{equation}
and similarly, $\p\lt[Y_{t+1}=0\rt]=1-p$. Under this simplifying
assumption, the preceding conditional transition matrix becomes
\begin{equation}
\begin{array}{c|c|c}
 & X_{t+1}(e)=0 & X_{t+1}(e)=1\\
\hline
X_t(e)=0 & \frac{\binom{n}{2}-m_{t}-p}{\binom{n}{2}-m_{t}} & \frac{p}{\binom{n}{2}-m_{t}} \\
\hline
X_t(e)=1 & \frac{1-p}{m_{t}} & \frac{m_{t}-1+p}{m_{t}} \\
\end{array}
\notag
\end{equation}
Each entry is obtained by taking the expectation with respect to $Y_{t}$. That is,
$\p[X_{t+1}(e)=\ome|X_{t}(e)=\ome\pri] =
\E[\p[X_{t+1}(e)=\ome|X_{t}(e)=\ome\pri,Y_{t}]]$, for every
$\ome,\ome\pri\in\{0,1\}$, and where the marginal distribution of $Y_{t}$
is known from equation (\ref{eq:independence}).

One can now combine the noise process described in section
\ref{sec:hmm}, with the birth and death stochastic process in order to
link the latent and observed parts of the Markov hidden model. This
gives the following table,
\begin{equation}
\begin{array}{c|c|c}
 & X\as_{t+1}(e)=0 & X\as_{t+1}(e)=1\\
\hline
X_t(e)=0 & (1-\alpha)\lt(\frac{\binom{n}{2}-m_{t}-p}{\binom{n}{2}-m_{t}}\rt) &
\alpha\lt(\frac{p}{\binom{n}{2}-m_{t}}\rt) \\
\hline
X_t(e)=1 & \beta\lt(\frac{1-p}{m_{t}}\rt) &
(1-\beta)\lt(\frac{m_{t}-1+p}{m_{t}}\rt) \\
\end{array}
\notag
\end{equation}
Since this transition matrix links the latent and observed stochastic
processes, one can immediately derive the marginal probabilities of the
$X\as_{t}(e)$'s, such that 
\begin{equation}
    \p[X\as_{t+1}(e)=0] =
    (1-\alpha)\lt(\frac{\binom{n}{2}-m_{t}-p}{\binom{n}{2}-m_{t}}\rt) 
    + \beta\lt(\frac{1-p}{m_{t}}\rt)\!,
\notag
\end{equation}
and similarly for $\p[X\as_{t}(e)=1]$.

The resulting $X_{t}$ process is \textit{non-stationary}.
Moreover, this also holds when considering the case $p=1-q$.
Indeed, since the probability of adding a new edge at time $t+1$ is
dependent on the number of existing edges,
$m_{t}:=|E_{t}|$, at time $t$; it follows that the resulting joint
distribution of any subset of the $X_{t}$'s  depends on the choice of $t$.

\small
\addcontentsline{toc}{section}{References}
\bibliographystyle{/home/cgineste/ref/style/oupced3}
\bibliography{/home/cgineste/ref/bibtex/Statistics,%
              /home/cgineste/ref/bibtex/Neuroscience}

\begin{thebibliography}{22}
\providecommand{\natexlab}[1]{#1}

\bibitem[{Achlioptas et~al.(2009)Achlioptas, D'Souza, and
  Spencer}]{Achlioptas2009}
Achlioptas, D., D'Souza, R.M., and Spencer, J. (2009).
\newblock Explosive percolation in random networks.
\newblock \textit{Science}, \textbf{323}(5920), 1453--1455.

\bibitem[{Alon and Spencer(2004)}]{Alon2004}
Alon, N. and Spencer, J. (2004).
\newblock \textit{The Probabilistic Method}.
\newblock Springer, London.

\bibitem[{Beveridge et~al.(2007)Beveridge, Bohman, Frieze, and
  Pikhurko}]{Beveridge2007}
Beveridge, A., Bohman, T., Frieze, A., and Pikhurko, O. (2007).
\newblock Product rule wins a competitive game.
\newblock \textit{Proc. Amer. Math. Soc}, \textbf{135}, 3061--3071.

\bibitem[{Bohman and Frieze(2001)}]{Bohman2001}
Bohman, T. and Frieze, A. (2001).
\newblock Avoiding a giant component.
\newblock \textit{Random Structures and Algorithms}, \textbf{19}, 75.

\bibitem[{Casteigts et~al.(2012)Casteigts, Flocchini, Quattrociocchi, and
  Santoro}]{Casteigts2012}
Casteigts, A., Flocchini, P., Quattrociocchi, W., and Santoro, N. (2012).
\newblock Time-varying graphs and dynamic networks.
\newblock \textit{International Journal of Parallel, Emergent and Distributed
  Systems}, \textbf{27}(5), 387--408.

\bibitem[{Ephraim and Merhav(2002)}]{Ephraim2002}
Ephraim, Y. and Merhav, N. (2002).
\newblock Hidden markov processes.
\newblock \textit{{IEEE} Transactions on Information Theory}, \textbf{48}(6),
  1518--1569.

\bibitem[{Erd\H{o}s and R\'{e}nyi(1960)}]{Erdos1960}
Erd\H{o}s, P. and R\'{e}nyi, A. (1960).
\newblock The evolution of random graphs.
\newblock \textit{Publ.~Math.~Inst.~Hungar.~Acad.~Sci.}, \textbf{5}, 17.

\bibitem[{Flocchini et~al.(2012)Flocchini, Mans, and Santoro}]{Flocchini2012}
Flocchini, P., Mans, B., and Santoro, N. (2012).
\newblock On the exploration of time-varying networks.
\newblock \textit{Theoretical Computer Science}, \textbf{469}, 53--68.

\bibitem[{Grindrod and Higham(2009)}]{Grindrod2009}
Grindrod, P. and Higham, D. (2009).
\newblock Evolving graphs: Dynamical models, inverse problems and propagation.
\newblock \textit{Proceedings of the {R}oyal {S}ociety, {S}eries {A}},
  \textbf{0456}, 1--18.

\bibitem[{Guye et~al.(2006)Guye, Regis, Tamura, Wendling, Gonigal, Chauvel, and
  Bartolomei}]{Guye2006}
Guye, M., Regis, J., Tamura, M., Wendling, F., Gonigal, A.M., Chauvel, P., and
  Bartolomei, F. (2006).
\newblock The role of corticothalamic coupling in human temporal lobe epilepsy.
\newblock \textit{Brain}, \textbf{129}(7), 1917--1928.

\bibitem[{Kramer et~al.(2010)Kramer, Eden, Kolaczyk, Zepeda, Eskandar, and
  Cash}]{Kramer2010}
Kramer, M., Eden, U., Kolaczyk, E., Zepeda, R., Eskandar, E., and Cash, S.
  (2010).
\newblock Coalescence and fragmentation of cortical networks during focal
  seizures.
\newblock \textit{Journal of Neuroscience}, \textbf{30}, 10076--10085.

\bibitem[{Kramer et~al.(2009)Kramer, Eden, Cash, and Kolaczyk}]{Kramer2009}
Kramer, M., Eden, U., Cash, S., and Kolaczyk, E. (2009).
\newblock Network inference with confidence from multivariate time series.
\newblock \textit{{P}hys. {R}ev. {E}.}, \textbf{79}, 061916.

\bibitem[{Kramer and Cash(2012)}]{Kramer2012}
Kramer, M.A. and Cash, S.S. (2012).
\newblock Epilepsy as a disorder of cortical network organization.
\newblock \textit{The Neuroscientist}, \textbf{18}(4), 360--372.

\bibitem[{Krivelevich et~al.(2010)Krivelevich, Lubetzky, and
  Sudakov}]{Krivelevich2010}
Krivelevich, M., Lubetzky, E., and Sudakov, B. (2010).
\newblock Hamiltonicity thresholds in {A}chlioptas processes.
\newblock \textit{Random Struct. Alg.}, \textbf{37}(1), 1--24.

\bibitem[{Lentz et~al.(2013)Lentz, Selhorst, and Sokolov}]{Lentz2013}
Lentz, H.H.K., Selhorst, T., and Sokolov, I.M. (2013).
\newblock Unfolding accessibility provides a macroscopic approach to temporal
  networks.
\newblock \textit{Phys. Rev. Lett.}, \textbf{110}(11), 118701.

\bibitem[{Ponten et~al.(2007)Ponten, Bartolomei, and Stam}]{Ponten2007}
Ponten, S., Bartolomei, F., and Stam, C. (2007).
\newblock Small-world networks and epilepsy: graph theoretical analysis of
  intracerebrally recorded mesial temporal lobe seizures.
\newblock \textit{Clinical neurophysiology}, \textbf{118}(4), 918--927.

\bibitem[{Riordan and Warnke(2012)}]{Riordan2012}
Riordan, O. and Warnke, L. (2012).
\newblock Achlioptas process phase transitions are continuous.
\newblock \textit{Annals of Applied Probability}, \textbf{4(22)}, 1450--1464.

\bibitem[{Riordan and Warnke(2011)}]{Riordan2011}
Riordan, O. and Warnke, L. (2011).
\newblock Explosive percolation is continuous.
\newblock \textit{Science}, \textbf{333}(6040), 322--324.

\bibitem[{Rubinov and Sporns(2010)}]{Rubinov2010}
Rubinov, M. and Sporns, O. (2010).
\newblock Complex network measures of brain connectivity: Uses and
  interpretations.
\newblock \textit{Neuroimage}, \textbf{52}, 1059--1069.

\bibitem[{Schindler et~al.(2010)Schindler, Amor, Gast, Muller, Stibal, Mariani,
  and Rummel}]{Schindler2010}
Schindler, K., Amor, F., Gast, H., Muller, M., Stibal, A., Mariani, L., and
  Rummel, C. (2010).
\newblock Peri-ictal correlation dynamics of high-frequency (80--200{Hz})
  intracranial {EEG}.
\newblock \textit{Epilepsy research}, \textbf{89}(1), 72--81.

\bibitem[{Schindler et~al.(2007{\natexlab{a}})Schindler, Elger, and
  Lehnertz}]{Schindler2007a}
Schindler, K., Elger, C.E., and Lehnertz, K. (2007{\natexlab{a}}).
\newblock Increasing synchronization may promote seizure termination: evidence
  from status epilepticus.
\newblock \textit{Clinical neurophysiology}, \textbf{118}(9), 1955--1968.

\bibitem[{Schindler et~al.(2007{\natexlab{b}})Schindler, Leung, Elger, and
  Lehnertz}]{Schindler2007}
Schindler, K., Leung, H., Elger, C.E., and Lehnertz, K. (2007{\natexlab{b}}).
\newblock Assessing seizure dynamics by analysing the correlation structure of
  multichannel intracranial {EEG}.
\newblock \textit{Brain}, \textbf{130}(1), 65--77.

\end{thebibliography}
\end{document}